\documentclass[apj]{emulateapj}
\slugcomment{{\sc Accepted for publication in the ApJ}}

\newcommand{\sgr}{SGR~1900+14}

\newcommand{\cxo}{{\it Chandra}}
\newcommand{\xmm}{{\it XMM-Newton}}
\usepackage{natbib}

\begin{document} 
\vspace{0.8 in} 

\title{Long Term Radiative Behavior of \sgr} 

\author{Ersin~{G\"o\u{g}\"u\c{s}}\altaffilmark{1},
Tolga~G\"uver\altaffilmark{2},
Feryal~\"Ozel\altaffilmark{2},
David~Eichler\altaffilmark{3},
Chryssa~Kouveliotou\altaffilmark{4}
}
 
\altaffiltext{1}{Sabanc\i~University, Faculty of Engineering and Natural Sciences, Orhanl\i-Tuzla, Istanbul 34956, Turkey}
\altaffiltext{2}{Department of Astronomy and Steward Observatory, University 
of Arizona, 933 N. Cherry Ave, Tucson, AZ, 85721, USA}
\altaffiltext{3}{Department of Physics, Ben-Gurion University of the 
Negev, Beer-Sheva 84105, Israel}
\altaffiltext{4}{Space Science Office, VP62, NASA/Marshall Space Flight Center, 
Huntsville, AL 35812, USA}

\begin{abstract} 

The prolific magnetar \sgr\ showed two outbursts in the last decade
and has been closely monitored in the X-rays to track the changes in
its radiative properties.  We use archival \cxo\ and \xmm\
observations of \sgr\ to construct a history of its spectrum and
persistent X-ray flux spanning a period of about seven years. We show
that the decline of its X-ray flux in these two outburst episodes
follows the same trend. The flux begins to decline promptly and
rapidly subsequent to the flares, then decreases gradually for about
600 days, at which point it resumes a more rapid decline. Utilizing
the high quality spectral data in each epoch, we also study the
spectral coevolution of the source with its flux. We find that neither
the magnetic field strength nor the magnetospheric properties change
over the period spanned by the observations, while the surface
temperature as well as the inferred emitting area both decline with
time following both outbursts.  We also show that the source reached
the same minimum flux level in its decline from these two subsequent
outbursts, suggesting that this flux level may be its steady quiescent
flux.

\end{abstract} 

\keywords{pulsars: individual (\sgr) $-$ X-rays: bursts}

\section{Introduction} 

Soft Gamma Repeaters (SGRs) and Anomalous X-ray Pulsars (AXPs) belong to a
class of objects called magnetars -- neutron stars whose X-ray
emission is likely to be powered by the decay of their extremely
strong magnetic fields (Duncan \& Thompson 1992; Thompson \& Duncan
1996; Thompson, Lyutikov \& Kulkarni 2002). All seven confirmed SGRs
and six out of seven confirmed AXPs\footnotemark{}\footnotetext{An
online catalog of general properties of SGRs and AXPs can be found at
http://www.physics.mcgill.ca/~pulsar/magnetar/main.html} have emitted
energetic bursts of X-rays/soft gamma rays. Burst active episodes of
magnetars last anywhere from few hours to months. During their
bursting activity, magnetars also exhibit remarkable temporal and
spectral changes in their persistent X-ray output. A detailed
description of SGRs and AXPs can be found in Woods \& Thompson (2006)
and Mereghetti (2008).

In the last seven years, new magnetar candidates have emerged, most
prominently through transient outbursts (e.g., XTE~J1810$-$197, 
Halpern \& Gotthelf 2005; CXO J164710.2-455216, Israel et al. 2007;
SGR J1833$-$0832, G\"o\u{g}\"u\c{s} et al. 2010). These sources were
too dim to be detected in X-rays during their quiescent phases
(usually below our detection sensitivity), but their X-ray fluxes
increased by up to few hundred times as they entered their outburst
episodes. In addition, known magnetar sources also exhibit variations
(triggered by bursting activity) in their persistent flux, although
typically not as dramatic as those in the transient systems (e.g.,
Woods et al. 2004). These high X-ray luminosities were instrumental in
probing the outburst mechanism of magnetars (\"Ozel \& G\"uver 2007;
G\"uver et al. 2007; Ng et al. 2010). In contrast, the low but
persistent flux level of most magnetars requires long term (5-10
years) monitoring to understand (burst-induced) changes in their
emission properties (Dib, Kaspi, \& Gavriil 2009).

\sgr\ has been one of the most prolific SGRs: it was discovered in 1979 
(Mazets, Golenetskij \& Guryan 1979), and was detected in a bursting
mode again in 1992 (Kouveliotou et al. 1993). In May 1998 the source
entered a major outburst episode that lasted about eight months and
included the giant flare on 1998 August 27 (Hurley et al. 1999). ASCA
and RXTE observations prior to and during the 1998 activation led to
the discovery of its 5.16~s spin period (Hurley et al. 1999), its
spin-down rate of $\sim$10$^{-11}$ s/s and magnetic field of
$2-8\times10^{14}$~G (Kouveliotou et al. 1999), and thus to the
confirmation of its magnetar nature. The source resumed a high level
of activity in April 2001 (Guidorzi et al. 2001; Kouveliotou et
al. 2001) and again in March 2006 (Vetere et al. 2006). As its burst
active phases are well separated from each other, \sgr\ is an
excellent source to investigate radiative changes both during bursting
behavior as well as in burst quiescence.

The first major enhancement in the persistent X-ray flux of \sgr\ was
observed at the onset of the 1998 August 27 giant flare: the flux
increased by a factor of $\sim$700 with respect to its level before
the activation (Woods et al. 2001). A detailed spectral analysis of
the (much longer) flux decay period was not possible due to the lack
of continuous monitoring observations with imaging instruments during
the decay phase. Esposito et al. (2007) analyzed nine pointed BeppoSAX
observations of \sgr\ spanning about five years from May 1997 to April
2002 and noted that the (2$-$10 keV) flux measured in the last
pointing faded significantly with respect to earlier
observations. Following the April 2001 activation, the \cxo\ X-ray
Observatory and \xmm\ observed \sgr\ at numerous occasions,
establishing a valuable dataset for understanding the long term
behavior of this source in, particular, and of magnetars, in general.

In this paper, we make use of all archival \cxo\ and \xmm\
observations to construct the persistent X-ray flux temporal and
spectral history of \sgr\ spanning about seven years following the
April 2001 activation. In the next section we introduce the \cxo\ and
\xmm\ observations used in this study. In Section 3, we present
the results of the spectral analysis and show that the magnetic field
strength and the magnetospheric properties remain stable over the
period spanned by the observations, while the surface temperature and
the inferred emitting area both decline with time following both
outbursts. We also show that the source flux shows the same trend in
its decline from outburst in both episodes and ultimately reaches the
same minimum flux in both cases.  We discuss the implications of these
results in Section 4.

\section{Observations and Data Analysis}

Between 2001 April 22 and 2008 April 8, \sgr\ was observed 13 times
with \cxo\ and \xmm\.  Table \ref{tbl:obss} lists the log
of these pointed X-ray observations. We describe below the details of
our data reduction.

\begin{table}[h]
\begin{center}
9
\caption{Log of \sgr\ Observations
\label{tbl:obss}}

\begin{tabular}{ccccc}
\hline
\hline

Observatory & Observation ID & Observation    & Exposure Time  \\
            &                &   Date         & (ks)           \\
\hline
\cxo\	    &	 2458	     &   2001 Apr 22  &  20.1      \\
\cxo\       &	 2459	     &   2001 Apr 30  &  18.8       \\
\cxo\       &	 3858	     &   2002 Nov 6  &  48.0      \\ 
\cxo\	    &	 3862	     &   2003 Feb 18  &  25.1      \\ 
\cxo\	    &	 3863	     &   2003 Jun 2  &  25.6      \\ 
\cxo\	    &	 3864	     &   2003 Oct 18  &  25.3      \\ 
\xmm\       &	 0305580101  &   2005 Sep 20  &  20.2      \\ 
\xmm\       &	 0305580201  &   2005 Sep 22  &  18.7      \\ 
\cxo\	    &	 6709	     &   2006 Mar 29  &  40.0      \\ 
\xmm\       &	 0410580101  &   2006 Apr 1  &  13.4      \\       
\cxo\	    &	 7593	     &   2007 Jun 24  &  12.2       \\
\cxo\	    &	 8215	     &   2007 Nov 21  &  12.6       \\
\xmm\       &	 0506430101  &   2008 Apr 8  &  18.7       \\
\hline
\end{tabular}
\end{center}
\end{table}

\subsection{{\it Chandra}}

There are a total of nine \cxo\ observations performed between 2001 April
22 and 2007 November 21. These were all performed using the Advanced
CCD Imaging Spectrometer (ACIS) in continuous clocking (CC) mode. We
selected rectangular source regions centered at the position of \sgr\
with dimensions $8 \arcsec \times 2 \arcsec$.  Our background regions
were selected with similar rectangular sizes from source-free regions
on the collapsed CC mode image. We calibrated the \cxo\ observations
using the CIAO version 4.2 and the CALDB version 4.2.

We extracted the X-ray spectra following the standard \cxo\ data
analysis threads with the psextract tool. We then used the mkacisrmf
and mkarf tools to create the detector and ancillary response files,
respectively. Finally, we re-binned the spectra such that each energy
bin would contain at least 50 counts, to decrease the formal errors.

\subsection{\xmm}

We analyzed the X-ray data obtained with the EPIC-pn detector between
2005 September 20 and 2008 April 8. In all observations, the EPIC-pn
was operating in full frame mode.  The calibration of the data was
performed with the Science Analysis Software (SAS) version 9.0.0 and
the latest available calibration files as of February 2010, using the
task epproc.

We extracted the X-ray source spectra by accumulating events from a
circular region with a radius of 32$\arcsec$ centered at \sgr. The
background regions were selected on the same chip from source free
regions with a typical radius of 50$\arcsec$. We then used the rmfgen
and arfgen tools of SAS to generate detector and ancillary response
files and re-binned the X-ray spectra such that each energy bin
contained at least 50 counts.

\section{Spectral Analysis and Results}

We used XSPEC v12.5.1n (Arnaud 1996) to analyze all spectra. We
assumed a gravitational redshift correction of 0.306, corresponding to
a neutron star mass of 1.4~$M_{\odot}$ and a radius of 10~km. We
performed the fits in the 0.8$-$6.5 keV range, where the lower energy
bound is set by the source flux and the higher energy limit is
determined by the non-thermal hard X-ray component (G\"otz et al.\
2006) affecting the soft X-ray spectra.

We fit the X-ray spectra using the Surface Thermal Emission and
Magnetospheric Scattering (STEMS; G\"uver, \"Ozel and Lyutikov 2006)
model. STEMS is based on the radiative equilibrium atmosphere
calculations presented in \"Ozel (2001, 2003) but also includes the
effects of magnetospheric scattering on the photons emitted from the
neutron star surface as calculated by Lyutikov \& Gavriil (2006). The
model parameters consist of the surface magnetic field strength, $B$,
surface temperature, $T$, the magnetospheric scattering optical depth,
$\tau$, and the velocity of the particles in the magnetosphere,
$\beta=v/c$, where {\it c} is the speed of light.

We first fit all 13 spectra simultaneously using STEMS and taking into
account the effect of interstellar absorption. We allowed the spectral
parameters of all spectra to vary individually. While we did not
specify its value a priori, we forced the hydrogen column density to
be the same between observations. We found that the magnetic field
strength, the scattering optical depth, and the average particle
velocity did not vary significantly between different observations. We
measured a hydrogen column density of $(2.36 \pm 0.05) \times
10^{22}$~cm$^{-2}$, assuming solar abundances (Anders \& Grevesse
1989). We, therefore, performed all further STEMS fits fixing the
hydrogen column density at the above value and forcing the parameters
{\it B}, $\tau$, and $\beta$ to be constant between different
observations. We still allowed the temperature and the model
normalization to vary individually for all spectra. We obtained good
fits to all 13 spectra, with a $\chi^2/$dof = 2370/2438. In this
combined fit, we found a magnetic field strength of $B = (5.0 \pm 0.3)
\times 10^{14}$~G, a scattering optical depth of $\tau = 8.7 \pm 0.9$
and a particle velocity of $\beta = 0.37 \pm 0.01$. Below, we discuss
the time evolution of the temperature $T$ and the source flux based on
these combined fits. Note that errors reported throughout the paper are 
1 $\sigma$.

\begin{figure}
\vspace{0.3in}
\centerline{
\includegraphics[angle=-90,width=3.2in]{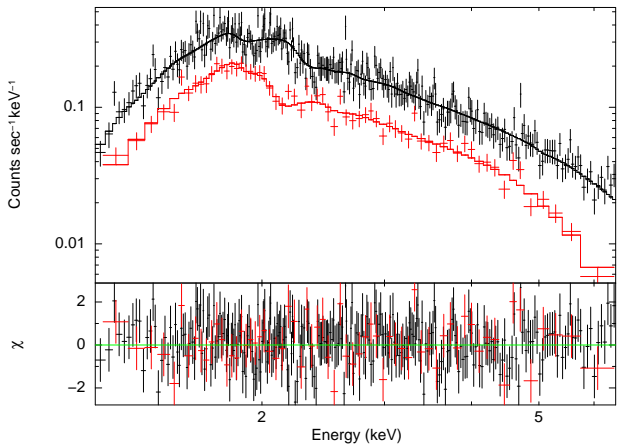}}
\caption{Upper panel: two X-ray spectra of \sgr\ observed with 
the \xmm\/EPIC-pn (upper data set in black) and the {\it
Chandra}/ACIS in CC mode (lower data set in red). Solid lines are the
best fitting STEMS model curves. Lower panel: residuals of the
spectral fits.
\label{fig:twospec}}
\end{figure}

We also investigated the potential effects of the cross calibration
between the {\it Chandra}/ACIS in CC mode and the {\it
XMM-Newton}/EPIC-pn on the determination of spectral parameters.
To this end, we used two data sets, observation ID 7593 with
\cxo\ and 0410580101 with \xmm\, which were selected because 
they occurred at comparable source fluxes.  We fit both spectra using
STEMS as well as the empirical blackbody plus power law model,
accounting for interstellar absorption in both cases. In this
analysis, we again forced the absorption parameter to remain constant
for both spectra as it is not observed to vary over time.  In
Figure~\ref{fig:twospec}, we present the two spectra along with the
best fitting STEMS models. We found all parameters of both continuum
models to be consistent between the {\it Chandra} and XMM spectra to
within 1 $\sigma$ errors. These results ensure that the use of two
different instruments does not introduce any systematic biases in the
joint spectral analysis.

\begin{figure}
\vspace{-0.0in}
\centerline{
\includegraphics[width=3.5in]{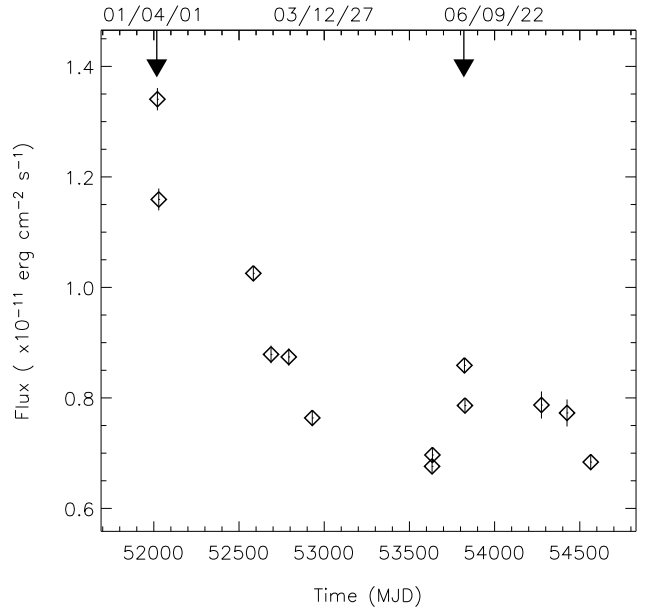}}
\vspace{0.1in}
\caption{Unabsorbed flux history of \sgr\ in the 0.8$-$6.5 keV range. 
The two arrows indicate the onset of the April 2001 and March 2006
reactivations of the source, respectively. The calendar dates on top
of the figure are in the YY/MM/DD format.
\label{fig:flxhist}}
\end{figure}

We present in Figure~\ref{fig:flxhist} the history of the unabsorbed
X-ray flux of \sgr\ in the 0.8$-$6.5~keV band. The first \cxo\ 
observation took place four days after the intermediate event (Feroci
et al.\ 2002). The source flux declined rapidly in the period soon
after this event, dropping from ($1.34 \pm 0.02) \times
10^{-11}$~erg~cm$^{-2}$~s$^{-1}$ to ($1.16 \pm 0.02) \times
10^{-11}$~erg~cm$^{-2}$~s$^{-1}$ in only about six days. An even faster
decline was seen in the contemporaneous {\it BeppoSAX} observations:
the source flux in the 2$-$10 keV band was ($3.1 \pm 0.3) \times
10^{-11}$~erg~cm$^{-2}$~s$^{-1}$ on 2001 April 18 and declined to
($1.06 \pm 0.04) \times 10^{-11}$~erg~cm$^{-2}$~s$^{-1}$ in 11 
days (Esposito et al. 2007). We observed a similar trend following the
2006 reactivation of the source as its flux drops from ($8.6 \pm 0.1)
\times 10^{-12}$~erg~cm$^{-2}$~s$^{-1}$ to ($7.9 \pm 0.1) \times
10^{-12}$~erg~cm$^{-2}$~s$^{-1}$ in three days.

\begin{figure}
\vspace{-0.2in}
\centerline{
\includegraphics[width=3.3in]{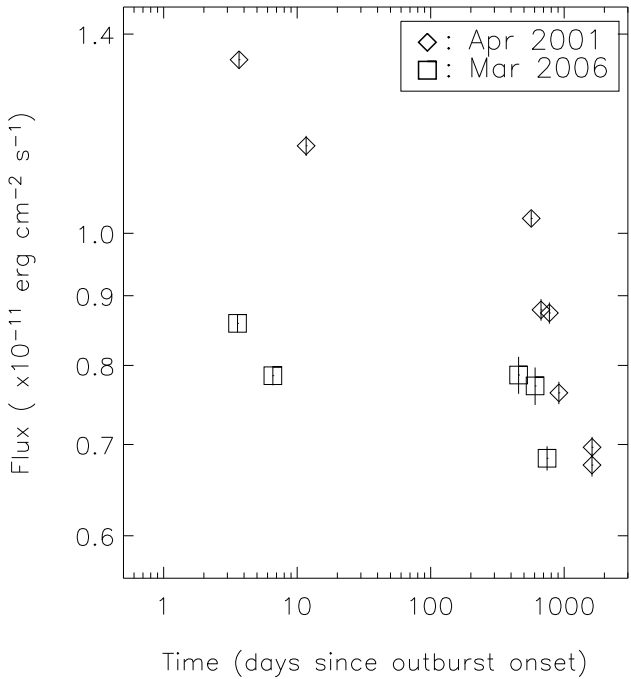}}
\vspace{0.2in}
\caption{Unabsorbed flux of \sgr\ in the $0.8-6.5$ keV range vs. relative 
time since the onset of the two bursting episodes: 2001 April
(triangles) and 2006 March (squares).
\label{fig:flxhistlog}}
\end{figure}

We further studied these long term persistent flux variations of \sgr\
as follows: we determined the relative times of the first eight
observations (which took place before 2006) with respect to the onset
of the April 2001 outburst and those of the remaining five
observations with respect to the onset of the March 2006
activation. We present in Figure~\ref{fig:flxhistlog} the unabsorbed
flux as a function of relative time since each respective outburst
onset. We found that in both outbursts the source flux declined
rather gradually until about 600~days after the onset and exhibited a
sharper decline trend beyond $\sim 600$~days. We also found that \sgr\
was at its lowest X-ray flux level of $6.8 \times
10^{-12}$~erg~cm$^{-2}$~s$^{-1}$ before the March 2006
reactivation. Following the source re-brightening after the 2006
outburst, the flux reached that level again in the last pointed
observation.

\begin{figure}
\vspace{-0.0in}
\centerline{
\includegraphics[width=3.3in]{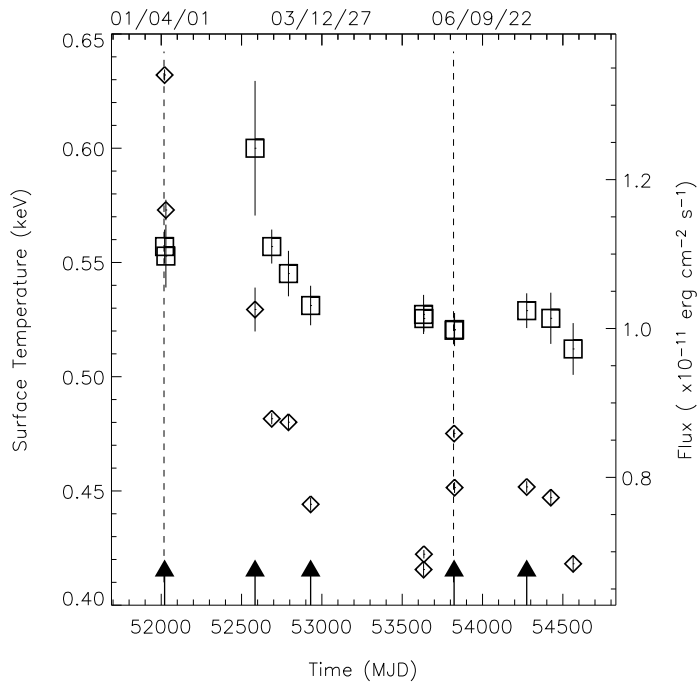}}
\vspace{0.0in}
\caption{Evolution of the neutron star surface temperatures (squares) 
obtained by fitting each spectrum with STEMS, and the corresponding
unabsorbed flux (diamonds). The vertical dashed lines indicate the
onsets of the 2001 and 2006 outbursts, respectively. Arrows at the
lower end of the figure indicate observations used in further phase
resolved spectral analysis (see the text). Calendar dates on top are
in the YY/MM/DD format.
\label{fig:flxtemp}}
\end{figure}

To better understand the nature of these flux variations in \sgr, we investigated a possible correlation between the flux and the only varying STEMS parameter, i.e., the surface temperature. 
We find that the flux and surface
temperature are indeed correlated, with a Spearman's rank order
correlation coefficient $r = 0.72$. The probability of obtaining such
a correlation with a random data set is $P = 0.005$. In
Figure~\ref{fig:flxtemp}, we show the history of the surface
temperature of the neutron star as well as its long term flux
behavior. The decline in the surface temperature alone (assuming a
single temperature) cannot account for the observed flux decline. In
particular, the average surface temperature dropped from $\sim
0.56$~keV to 0.53~keV as measured on 2001 April 22 (MJD 52021) and
2005 September 22 (MJD 53635), respectively, which corresponds to a
flux decline of about 25$\%$ (if the emitting surface area remains
constant). However, the source flux declined by about 97$\%$ between
these two epochs, mostly soon after the rise; the remaining decline
over the last 1100 days of the observing period was only about 33$\%$,
while the temperature decreased by about 6$\%$.

We also investigated the long term behavior of the surface emitting
area inferred using the STEMS normalization and surface
temperature. We found that, for a uniform surface temperature, the
radius of the emitting region is 4.8$\pm$0.2~km soon after the 2001
outburst onset (assuming a distance to \sgr\ of 13.5~kpc; Vrba et
al. 2000). It then declined to 3.9$\pm$0.2~km and remained fairly
constant until 2005. The radius emitting surface went up again to
4.4$\pm$0.1~km following the 2006 outburst but quickly fell to a
marginally higher constant level of 4.1$\pm$0.2 km. This change,
however, could be the result of neglecting temperature
inhomogeneities. Note also that the errors in the surface temperature
and normalization, and, consequently, in the radius of the emitting
area, are correlated.

\begin{figure}
\vspace{-0.0in}
\centerline{
\includegraphics[width=3.3in]{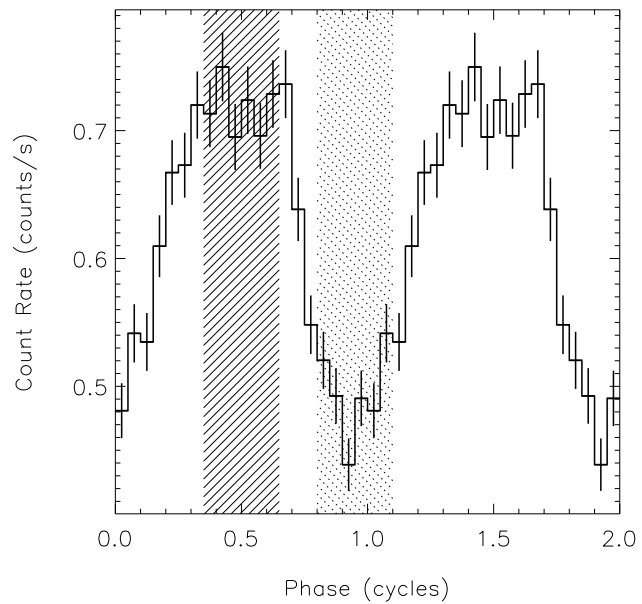}}
\vspace{0.1in}
\caption{Pulse profile of \sgr\ during the \cxo\ observations on 
2001 April 22 (MJD 52021). The pulse phase intervals used as pulse
peak and pulse minimum spectral accumulations are shown with hatched
and dotted regions, respectively.
\label{fig:phaseres}}
\end{figure}

Finally, we performed a coarse phase resolved spectral analysis using
five \cxo\ observations (indicated by arrows in Figure
\ref{fig:flxtemp}) to check whether there are significant surface
temperature variations over the spin phase, which may affect or bias
the spectral determination of the phase-averaged temperature.  For
each of the selected pointings, we obtained a pulse peak spectrum
(spanning 0.30 of the spin phase during the pulse peak) and a pulse
minimum spectrum (spanning 0.30 of the spin phase during the pulse
minimum) using accurate and contemporaneous pulse period reported in 
Mereghetti et al. (2006). We show in Figure~\ref{fig:phaseres} the pulse 
profile and pulse phase intervals within which the pulse peak and pulse minimum
spectra were obtained in the observation on 2001 April 22 (MJD
52021). We fit the X-ray spectra in the 0.8$-$6.5 keV using STEMS with
surface magnetic field strength and magnetospheric parameters fixed at
their phase averaged values. In Table \ref{tbl:cxo_specfit}, we list
the temperatures and model normalizations (i.e., an indicator of
intensity) for the five selected observations. We find that the
surface temperature remains constant over the spin phase of the
source in the majority of these observations.

\begin{table*}
\begin{center}

\caption{Spectral Fit Results of the Phase-resolved Analysis
\label{tbl:cxo_specfit}}

\begin{tabular}{ccccc}
\hline
\hline

Obs Date & Pulse Peak      & Pulse Minimum & Peak Normalization  & Min Normalization \\
(MJD)    & Temperature (keV) & Temperature (keV) & (10$^{-10}$)  & (10$^{-10}$) \\
\hline
52021    & 0.57$\pm$0.01     & 0.54$\pm$0.01     & 2.2$\pm$0.3   & 1.8$\pm$0.1  \\
52584    & 0.56$\pm$0.01     & 0.55$\pm$0.01     & 1.6$\pm$0.1   & 1.2$\pm$0.1  \\
52930    & 0.53$\pm$0.01     & 0.52$\pm$0.01     & 0.57$\pm$0.05 & 0.48$\pm$0.05 \\
\hline
53823    & 0.53$\pm$0.01     & 0.52$\pm$0.01     & 0.26$\pm$0.03 & 0.19$\pm$0.02 \\
54275    & 0.55$\pm$0.01     & 0.47$\pm$0.01     & 0.37$\pm$0.03 & 0.5$\pm$0.2  \\
\hline

\end{tabular}
\end{center}
\end{table*}

\vspace{1.0in}

\section{Discussion}

We found that \sgr\ exhibits a monotonic but non-steady flux decline
following the X-ray brightening during its 2001 and 2006 outbursts. In
both outbursts, its flux dropped rapidly within a few weeks after the
onset of the outburst and then at a slower rate for approximately
600~days. After this period, the flux declines again at a much faster
rate. A similar decay trend was also seen in SGR~1627$-$41
(Kouveliotou et al.\ 2003). The fact that we observe the same trend in
successive outbursts from the same source suggests that the long-term
effects of outbursts are not stochastic but reproducible.

It is clear that magnetar bursting activity leads to long-term flux
enhancements and that the additional energy powering these
enhancements is stored in the crust as heat. This heat comes most
likely from the energy released in the crust during the bursts, or
from the energy deposited in the crust by the bombardment with
magnetospheric particles during the burst. The crust, then, reradiates
this additional heat over a timescale of a few years.

Despite the correlation between the flux and the surface temperature,
the variation in the surface temperature alone does not account for
the decline in the flux, but it also requires a change in the emitting
area over time. This can perhaps be understood if the crust is heated
inhomogeneously, as would be expected if the initial heating episode
is due to magnetic energy release. Moreover, because the thermal
resistance of the crust is dominated by the uppermost layers, where
the heat conductivity is strongly affected by the magnetic field, heat
coming from the deeper layers of the crust could reach the surface
unevenly. This leads to both uneven heating and uneven cooling, which
may affect the total inferred emitting area. Furthermore, over-time,
because the observations are carried out in a limited energy range,
cooler parts of the crust may fall out of the observed energy band
faster than the hotter regions, again reducing the inferred emitting
area. Thus, the observed source flux would decline both due to a
decline in temperature as $\propto T^4$ and due to a decline in the
emitting area.

We find on three different occasions that the X-ray flux of \sgr\ was
as low as $6.8 \times 10^{-12}$~erg~cm$^{-2}$~s$^{-1}$, suggesting
that this level may correspond to the source persistent X-ray flux in
the absence of burst induced enhancements. The current detection
thresholds of imaging instruments are $\sim$10$^{-13}$ erg cm$^{-2}$
s$^{-1}$; it is, therefore, possible that the persistent flux levels
of the so-called transient magnetars are much lower than those of
the always detectable magnetars. If that is indeed the case, the
magnetar engine that is responsible for a source persistent quiescent
X-ray emission would seem to power a broad flux range of $\lesssim
10^{-13}$ to $\sim10^{-10}$~erg~cm$^{-2}$~s$^{-1}$.

\acknowledgments 

E.G., T.G., and F.\"O. acknowledge EU FP6 Transfer of Knowledge Project
Astrophysics of Neutron Stars (MTKD-CT-2006-042722). D.E. acknowledges
support from the Israel Science Foundation, the U.S. Israel Binational
Science Foundation, and the Joan and Robert Arnow Chair of Theoretical
Astrophysics.

\end{document}